\documentclass [smallextended]{svjour3}
\usepackage[T1]{fontenc}
\usepackage{geometry}
\usepackage{color}
\usepackage[english]{babel}
\usepackage{amsmath}
\usepackage{amssymb}
\usepackage{graphicx}
\usepackage[unicode=true,pdfusetitle,bookmarks=true,bookmarksnumbered=false,bookmarksopen=false,breaklinks=false,pdfborder={0 0 0},backref=false,colorlinks=true] {hyperref}
\hypersetup{linkcolor=blue, citecolor=blue, urlcolor=blue}


\usepackage[T1]{fontenc}

\usepackage{slashed}

\renewcommand{\emph}[1]{\textit{#1}}
\newcommand{\Tr}{\mathrm{Tr}}
\newcommand{\logd}{\log_{2}}

\AtBeginDocument{
  
}

\makeatother

\begin{document}

\title{Analytical Expression of Genuine Tripartite Quantum Discord for Symmetrical
X-states}

\author{Andrea Beggi \and Fabrizio Buscemi \and Paolo Bordone}

\institute{Andrea Beggi \and Fabrizio Buscemi \and Paolo Bordone 
\at Dipartimento di Scienze Fisiche, Informatiche e Matematiche,\\
Universit\`{a} di Modena e Reggio Emilia,\\
Via Campi 213/A, I-41125 Modena, Italy,\\\email{andrea.beggi@unimore.it}
\and
Paolo Bordone 
\at Centro S3, CNR - Istituto Nanoscienze,\\
Via Campi 213/A, I-41125 Modena, Italy,
}

\date{Received: date / Accepted: date}

\maketitle

\begin{abstract}
The study of classical and quantum correlations in bipartite and multipartite
systems is crucial for the development of quantum information theory.
Among the quantifiers adopted in tripartite systems, the genuine tripartite
quantum discord (GTQD), estimating the amount of quantum correlations
shared among all the subsystems, plays a key role since it represents
the natural extension of quantum discord used in bipartite systems.
In this paper, we derive an analytical expression of GTQD for three-qubit
systems characterized by a subclass of  symmetrical X-states.
Our approach has been tested on both GHZ and maximally mixed states
reproducing the expected results. Furthermore, we believe that the
procedure here developed constitutes a valid guideline to investigate
quantum correlations in form of discord in more general multipartite
systems.
\end{abstract}


\section{INTRODUCTION}

Quantum correlations are assuming increasing relevance, since they
can be exploited to improve our ability to perform many informational
and computational tasks~\cite{SEPARAB_ENTGNL_1QUBIT,SEPARAB_ENTGN_CORR,SEPAR_QDISC_1QUBIT,SEPAR_QUANT_COMP_WO_ENTNGL}.
Therefore, the problem of their characterization and quantification
has become a significant topic of research. Traditionally, the most
used form of quantum correlation is \emph{entanglement}, and the development
of quantum information theory is fundamentally due to its implementation
in information and communication protocols~\cite{NIELSEN_CHUANG,HORODECKI_REVIEW_ENTANGLEMENT}.

A form of quantum correlation other than entanglement is \emph{quantum
discord} (QD)~\cite{ZUR_DISC,H_VEDR,LUO_QBIT}, which can be expressed
in terms of the difference between the total and the classical correlations
for a system when one of its subparties is subject to an unobserved
measure process. Such a quantity, however, significantly depends upon
both the subsystem chosen and the measurement performed on it: in
particular, if the measurement is carefully selected, we can minimize
its ``disturbing effect'' on the system~\cite{LUO_MEAS}. This
choice corresponds to the minimization of QD firstly over a set of
possible measurements (on a fixed subsystem), typically \emph{projective
von Neumann measurements}~\cite{LUO_MEAS,ZUR_DISC}, and secondly
over all possible subsystem on which the local measurement can be
performed. Recent efforts in the study of the optimization processes
have led to analytical expression for quantum discord in some particular~\cite{LUO_QBIT} and more general states~\cite{GIROLAMI_ANAL_PROGR_DISC_2Q,QD_ARB_ST2} in systems composed
of two qubits.

Both entanglement and QD have widely been analyzed and used in bipartite
systems, while their extension to multipartite systems is still discussed
and tackled with different approaches~\cite{BENNET_MULTIP_POSTUL,RULLI_MULTIP_GLOBAL_QDISC,Xu_GLOBAL_QDISC_ANAL,CHAKRA_QDISS_3QBIT,GIORGI_GENUIN_MULTIP}.
For instance, \emph{Vinjanampathy et al.}~\cite{Vinja_Rau_QBIT-QDIT}
proposed a method to evaluate analytically quantum discord for a \emph{n}-partite
system of qubits in some special cases, but they treated the whole
system as a bipartite one (each subparty containing $1$ or $n-1$
qubits, respectively). On the other hand, \emph{Giorgi et al.}~\cite{GIORGI_GENUIN_MULTIP}
defined, for a \emph{n}-partite system, \emph{genuine n}-partite correlations\emph{,}
which can be divided into total, classical or quantum. These kinds
of correlations are shared between all the \emph{n} parties which
form the system, i.e. they cannot be accounted for considering any
of the possible subsystems. The quantum part of genuine correlations
is quantified by \emph{genuine n-partite quantum discord}.
The approach of Ref.~\cite{GIORGI_GENUIN_MULTIP} represents a natural
extension of the concept of QD as introduced for bipartite systems,
and this is the reason why we will follow it in the present work.
However, it requires massive numerical optimization procedures over
a number of parameters, thus making the calculations very demanding~\cite{Yichen_QD_NPCOMPLETE}.
Therefore, the application of such a criterion is not easily amenable.

This justifies the scarce number of works investigating the time evolution
of quantum correlations in multipartite system coupled to noisy environments.
Specifically only few cases have been considered: two level systems
undergoing random telegraph noise~\cite{FABRIZIO_GENUIN,THEO_GEN_MULTIP_CORR}
and quantum phase transitions in spin systems~\cite{EXP_MULTIP_OPSYS,EXP_MULTIP_SPINCHAIN,EXP_MULTIP_XXZModel}.

The purpose of this paper is to derive an analytical expression for
the \emph{genuine tripartite quantum discord} (GTQD)\emph{ }for a
class of three qubits systems.\emph{ }In detail, we will focus on
those systems described by X-states, which play a relevant role in
a large number of physical systems and allow for easy calculations
of certain entanglement measures~\cite{WEINST_TRIP_ENTANGL,WEINSTEIN_ENTNGL_3Q_X}.
X-states have been widely investigated also in bipartite systems, where an
analytical expression for QD has been proposed in \cite{ANAL_QD_2QUBIT_X_states}. However,
this approach has been questioned, since it is not always providing the correct result~\cite{Lu_counterxample_Ali_X_states_QD,Chen_Zang_QD_2qubit_Xstates,Yichen_QD_X_states_Symm_ERRORS}.

The paper is organized as follows. In Sec.~\ref{sec:QUANTIFIERS-FOR-GENUINE}
we introduce the genuine quantifiers for correlations in multipartite
quantum systems. In Sec.~\ref{sec:THREE-QUBITS-SYMMETRICAL-X-STATE}
we introduce the expression for a symmetrical tripartite X-state and
derive some constraints on its defining parameters. In Sec.~\ref{sec:ESTIMATION-OF-GENUINE}
we estimate all the quantities required to compute GTQD, and in particular
we describe the optimization procedures (both numerical and analytical)
appearing in the expression of GTQD. Sec.~\ref{sec:RESULTS-AND-DISCUSSION}
concerns the comparison between our results on GTQD and others already
present in the literature and, finally, in Sec.~\ref{sec:CONCLUSIONS}
we draw conclusions. 


\section{QUANTIFIERS FOR GENUINE TRIPARTITE CORRELATIONS\label{sec:QUANTIFIERS-FOR-GENUINE}}

Here we illustrate the correlation measures adopted in this work to
quantify tripartite quantum discord and entanglement.

\subsection{Tripartite Quantum Discord}

In a tripartite system, described by a state $\rho=\rho_{A,B,C}$,
the tripartite\emph{ }quantum mutual information\emph{ }is obtained
as a generalization of the quantum mutual information for bipartite
systems \cite{Vedr_REL_ENTR,Modi_UNIFIED_QUANT_CLASS_CORR,GIORGI_GENUIN_MULTIP,WALCZAK_MULTIPARTITE}:
\begin{equation}
T(\rho)=S(\rho_{A})+S(\rho_{B})+S(\rho_{C})-S(\rho),
\end{equation}

and represents the total amount of correlations encoded in this system\footnote{It can be shown that this quantity measures,  in terms of relative entropy, the distance between the state $\rho$  and the nearest classical state with no correlations $\rho^{A}\otimes\rho^{B}\otimes\rho^{C}$. Indeed, by the definition of relative entropy, we get $S(\rho||\rho^{A}\otimes\rho^{B}\otimes\rho^{C})=-\mathrm{Tr}[\rho\logd(\rho^{A}\otimes\rho^{B}\otimes\rho^{C})]-S(\rho)$, then using the linearity of trace and the additivity of logarithm - remember that $\rho^{i}$ are the marginals of $\rho$ - we get $-\mathrm{Tr}[\rho\logd(\rho^{A}\otimes\rho^{B}\otimes\rho^{C})]=-Tr[\rho\logd(\rho^{A})\otimes I\otimes I]+...=S(\rho^{A})+S(\rho^{B})+S(\rho^{C})$ \cite{Modi_UNIFIED_QUANT_CLASS_CORR,GIORGI_GENUIN_MULTIP}.}.
Here $S(\rho)=-\mathrm{Tr}[\rho\logd(\rho)]$ is the von Neumann entropy,
and $\rho_{i}$ $(i=A,B,C)$ is the reduced density matrix for the
subsystem $i$. Following the same procedure used in the literature
for bipartite systems~\cite{ZUR_DISC}, \emph{Giorgi et al.}~\cite{GIORGI_GENUIN_MULTIP}
define the \emph{tripartite classical correlations} in the system
as the quantum version (of a classical analogue) of the mutual information
derived from the Bayes' rule:
\begin{equation}
J(\rho)=\underset{i,j,k\in\{A,B,C\}}{\max}[S(\rho_{i})-S(\rho_{i|j})+S(\rho_{k})-S(\rho_{k|ij})],
\end{equation}

which has been optimized over the indices $i,j,k$ in the set of all
the possible permutation of subsystems $\{A,B,C\}$. Here $S(\rho_{i|j})$
and $S(\rho_{k|ij})$ are relative entropies and $\rho_{i|j}$ and
$\rho_{k|ij}$ are the density matrices after a measurement on the
subsystem $i$ or after a measurement on both subsystems $i$ and
$j$, respectively~\cite{FABRIZIO_GENUIN}. We refer the reader to
the Appendix~\ref{sec:Relative-Entropies-Definition} for a detailed
definition of the relative entropies and their optimization. Like
for bipartite systems, the tripartite quantum discord is given by
the difference between total and classical correlations:
\begin{equation}
D(\rho)=T(\rho)-J(\rho).
\end{equation}

However, among the correlations included in $T(\rho)$, a subset is
shared by all of the three subsystems (\emph{genuine tripartite mutual
information}), and can be estimated as:
\begin{equation}
T^{(3)}(\rho)=T(\rho)-T^{(2)}(\rho),\label{eq:T3}
\end{equation}

where $T^{(2)}(\rho)$ is the maximum amount of mutual information
shared by any couple of subsystems:
\begin{equation}
T^{(2)}(\rho)=\max[I(\rho_{A,B}),I(\rho_{A,C}),I(\rho_{B,C})],
\end{equation}

where $I(\rho_{AB})=S(\rho_{A})+S(\rho_{B})-S(\rho_{AB})$. Since
all the correlations that cannot be accounted for by $T^{(2)}(\rho)$
must be shared between all of the three subsystems, we can conclude
that $T^{(3)}(\rho)$ measures the distance between $\rho$ and the
closest product state along any bipartite cut of the system. Indeed
it can be shown that $T^{(3)}(\rho)=\min[I(\rho_{AB,C}),I(\rho_{AC,B}),I(\rho_{BC,A})]$
(see Ref.~\cite{GIORGI_GENUIN_MULTIP}).

Analogously, the genuine tripartite classical correlations reads:
\begin{equation}
J^{(3)}(\rho)=J(\rho)-J^{(2)}(\rho),\label{eq:J3}
\end{equation}

and GTQD:
\begin{equation}
D^{(3)}(\rho)=D(\rho)-D^{(2)}(\rho),\label{eq:D3}
\end{equation}

where%
\footnote{In Eqs. (\ref{eq:J2}) and (\ref{eq:D2}) we used the bipartite quantifiers
$J(\rho_{A,B})=\max[S(\rho_{A,B})-S(\rho_{A|B}),S(\rho_{A,B})-S(\rho_{B|A})]$
and $D(\rho_{A,B})=I(\rho_{A,B})-J(\rho_{A,B})$ as they are usually
defined in literature for 2-qubits systems \cite{GIORGI_GENUIN_MULTIP,FABRIZIO_GENUIN,LUO_QBIT}.%
}:
\begin{align}
J^{(2)}(\rho)=\max[J(\rho_{A,B}),J(\rho_{A,C}),J(\rho_{B,C})],\label{eq:J2}\\
D^{(2)}(\rho)=\min[D(\rho_{A,B}),D(\rho_{A,C}),D(\rho_{B,C})].\label{eq:D2}
\end{align}

Eqs. (\ref{eq:T3}), (\ref{eq:J3}) and (\ref{eq:D3}) can be significantly
simplified for the case of a state $\rho$ symmetrical under any exchange
of its subsystems. Indeed it can be shown that~\cite{FABRIZIO_GENUIN}:
\begin{align}
T^{(3)}(\rho)=I(\rho_{A,BC})=S(\rho_{A})+S(\rho_{A,B})-S(\rho),\\
D^{(3)}(\rho)=S(\rho_{A|BC})+S(\rho_{A,B})-S(\rho),\label{eq:GenTripDisc}\\
J^{(3)}(\rho)=S(\rho_{A})-S(\rho_{A|BC}).
\end{align}

\subsection{Tripartite Negativity}

In tripartite systems, represented by a state $\rho$, we can detect
the presence of entanglement between subsystems by using the \emph{negativity}
$N$, which is defined as follows~\cite{Vidal_Negativity}:
\begin{equation}
N(\rho^{tC})=\sum_{i}|\lambda_{i}(\rho^{tC})|-1.
\end{equation}

In the previous expression, $\rho^{tC}$ is the partial transpose
of $\rho$ with respect to the subsystem $C$, and $\lambda_{i}(\rho^{tC})$
are the eigenvalues of $\rho^{tC}$. The negativity can be equivalently
interpreted as the sum of the absolute values of the negative eigenvalues
of $\rho^{tC}$~\cite{Vidal_Negativity}, and it depends upon the
subsystem on which we make the partial transpose of $\rho$.

When negativity is higher than zero, we can conclude that there is
an entanglement between the subsystem $C$ and the compound subsystem
$A-B$, but the converse is not necessarily true. Starting from this
point, we can define the \emph{tripartite negativity} as follows~\cite{Sabin_Trip_Negat}:
\begin{equation}
N^{(3)}(\rho)=\sqrt[3]{N(\rho^{tA})N(\rho^{tB})N(\rho^{tC})},
\end{equation}

and this quantifier will be different from zero only when the entanglement
is shared among all of the three subsystems, i.e. it is a \emph{``full''
}tripartite entanglement~\cite{Sabin_Trip_Negat}. However, apart
from pure states, a null negativity could indeed not imply the absence
of entanglement. Moreover, we must notice that tripartite negativity
cannot distinguish the entanglement of a genuine tripartite entangled
state from that of a biseparable state in a generalized sense~\cite{Sabin_Trip_Negat,EXP_MULTIP_SPINCHAIN}.
For tripartite systems that are symmetrical under any exchange of
their qubits, as in our case of study, the tripartite negativity and
the negativity always coincide:
\begin{equation}
N^{(3)}(\rho)=N(\rho^{tA})=N(\rho^{tB})=N(\rho^{tC}).\label{eq:trip_neg}
\end{equation}

Another possible quantifier for tripartite entanglement is the \emph{three-tangle}~\cite{COFFMAN_Three_Tangle}, but in this work we use negativity
since the three-tangle is not able to detect tripartite entanglement
for all states, e.g. W states~\cite{Dur_Vidal_W_and_GHZ}.
However, it should be noticed that $N^{(3)}$ in this work is used simply to provide a further comparison with the outcomes of GTQD, and it is not used to quantify genuine entanglement.

\section{THREE-QUBITS SYMMETRICAL X-STATES\label{sec:THREE-QUBITS-SYMMETRICAL-X-STATE}}

Here, we focus on three qubits X-states~\cite{YU_X_States_intro}
which, for the particular features of their quantum correlations,
have been investigated in the literature, both for bipartite~\cite{Vinja_Rau_QBIT-QDIT,Chen_Zang_QD_2qubit_Xstates,ANAL_QD_2QUBIT_X_states} and tripartite systems~\cite{WEINSTEIN_ENTNGL_3Q_X,WEINST_TRIP_ENTANGL,FABRIZIO_GENUIN}.
A generic tripartite X-state can be written in the form~\cite{WEINSTEIN_ENTNGL_3Q_X}:
\begin{equation}
\rho=\left(\begin{array}{cccccccc}
a_{1} & 0 & 0 & 0 & 0 & 0 & 0 & c_{1}\\
0 & a_{2} & 0 & 0 & 0 & 0 & c_{2} & 0\\
0 & 0 & a_{3} & 0 & 0 & c_{3} & 0 & 0\\
0 & 0 & 0 & a_{4} & c_{4} & 0 & 0 & 0\\
0 & 0 & 0 & c_{4}^{*} & b_{4} & 0 & 0 & 0\\
0 & 0 & c_{3}^{*} & 0 & 0 & b_{3} & 0 & 0\\
0 & c_{2}^{*} & 0 & 0 & 0 & 0 & b_{2} & 0\\
c_{1}^{*} & 0 & 0 & 0 & 0 & 0 & 0 & b_{1}
\end{array}\right).
\end{equation}

In order to simplify the derivation of an analytical expression for
GTQD, we limit ourselves to X-states which are symmetrical under any
exchange of their subsystems, and invariant under the flip of all
of their qubits. This means that $\rho$ can be written in the form:

\begin{equation}
\rho=\frac{1}{8}{\scriptstyle \left(\begin{array}{cccccccc}
1-a_{1} & 0 & 0 & 0 & 0 & 0 & 0 & c_{1}\\
0 & \alpha_{1} & 0 & 0 & 0 & 0 & c_{2} & 0\\
0 & 0 & \alpha_{1} & 0 & 0 & c_{2} & 0 & 0\\
0 & 0 & 0 & \alpha_{1} & c_{2} & 0 & 0 & 0\\
0 & 0 & 0 & c_{2} & \alpha_{1} & 0 & 0 & 0\\
0 & 0 & c_{2} & 0 & 0 & \alpha_{1} & 0 & 0\\
0 & c_{2} & 0 & 0 & 0 & 0 & \alpha_{1} & 0\\
c_{1} & 0 & 0 & 0 & 0 & 0 & 0 & 1-a_{1}
\end{array}\right)},\label{eq:Rho_X}
\end{equation}

where $\alpha_{1}=1+\frac{a_{1}}{3}$ (we used the property $\mathrm{Tr}[\rho]=1$
to express $a_{2}$ in terms of $a_{1}$, and then we made the substitutions
$a_{1}\rightarrow\frac{1-a_{1}}{8}$, $c_{i}\rightarrow\frac{c_{i}}{8}$
to get a simpler expression). Now $\rho$ depends only on the parameters
$(a_{1},c_{1},c_{2})$ which, from now on, are
assumed to be real due to the qubit-flip invariance.
Recently, symmetry features of mixed entangled states have been also exploited
in Ref.~\cite{CAMPBELL_GD_SYMM_MIX} to evaluate analytically both nonlocality and global quantum discord
in multipartite systems.

From the requirement $0\leq\lambda_{i}\leq1\;\forall i$, where $\lambda_{1,2}=\frac{1}{8}\left(1-a_{1}\mp c_{1}\right)$
and $\lambda_{3-4-5,6-7-8}=\frac{1}{24}\left(3+a_{1}\mp3c_{2}\right)$
are the eigenvalues of $\rho$, we obtain the following constraints
for the parameters:
\begin{align}
a_{1} & \in\left[-3,1\right],\nonumber \\
c_{1} & \in\left[a_{1}-1,1-a_{1}\right]\label{eq:Constra1}\\
c_{2} & \in\left[-1-\frac{a_{1}}{3},1+\frac{a_{1}}{3}\right].\nonumber 
\end{align}

\section{ESTIMATION OF GENUINE TRIPARTITE QUANTUM DISCORD\label{sec:ESTIMATION-OF-GENUINE}}

\subsection{von Neumann Entropies for $\rho$ and $\rho_{A,B}$ }

Now, in order to give an analytical estimation of $D^{(3)}(\rho)$
for the state $\rho$ described by Eq.~(\ref{eq:Rho_X}), we calculate
the von Neumann entropies for $\rho$ and for the marginal $\rho_{A,B}=\Tr_{C}[\rho]$,
which appears in the expression of GTQD given by Eq.~(\ref{eq:GenTripDisc}).

From the definition of von Neumann entropy, it follows that: 
\begin{align}
S(\rho) & =3+\frac{1}{8}\left[2(3+a_{1})\log_{2}(3)-\left(1-a_{1}-c_{1}\right)\logd\left(1-a_{1}-c_{1}\right)-\left(1-a_{1}+c_{1}\right)\logd\left(1-a_{1}+c_{1}\right)\right.\nonumber \\
 & \left.-\left(3+a_{1}-3c_{2}\right)\logd\left(3+a_{1}-3c_{2}\right)-\left(3+a_{1}+3c_{2}\right)\logd\left(3+a_{1}+3c_{2}\right)\right].\label{eq:S_tot}
\end{align}

From Eq.~(\ref{eq:Rho_X}) we obtain:
\begin{equation}
\rho_{A,B}=\left(\begin{array}{cccc}
\frac{3-a_{1}}{12} & 0 & 0 & 0\\
0 & \frac{3+a_{1}}{12} & 0 & 0\\
0 & 0 & \frac{3+a_{1}}{12} & 0\\
0 & 0 & 0 & \frac{3-a_{1}}{12}
\end{array}\right),
\end{equation}

and after straightforward calculations we find:
\begin{align}
S(\rho_{A,B}) & =-\frac{1}{6}\left(3-a_{1}\right)\logd\left(3-a_{1}\right)-\frac{1}{6}\left(3+a_{1}\right)\logd\left(3+a_{1}\right)+2+\logd(3).\label{eq:S_AB}
\end{align}

\subsection{Relative entropy minimization}\label{sec:Rel_Entr_Minim}

In order to finally evaluate $D^{(3)}(\rho)$ we need to calculate
the relative entropy $S(\rho_{A|BC})$. Following the derivation procedure
given in the Appendix~\ref{sec:Relative-Entropies-Definition}, $S(\rho_{A|BC})$
can be written as:

\begin{align}
S(\rho_{A|BC})=\underset{\theta_{i},\phi_{i}}{\min}\, S_{rel}(\theta_{1},\theta_{2},\phi_{1},\phi_{2})=\underset{\theta_{i},\phi_{i}}{\min}\left\{ 1+\frac{1}{6}\left[\lambda_{A}\logd\lambda_{A}+\lambda_{B}\logd\lambda_{B}\right]-\frac{1}{12}\sum_{i=1}^{4}\lambda_{i}\logd\lambda_{i}\right\} ,\label{eq:SA|BC}
\end{align}

where $\theta_{i}$ and $\phi_{i}$ are optimization parameters (the
angles defining the basis vectors: see again Appendix~\ref{sec:Relative-Entropies-Definition}),
and:
\begin{align}
\lambda_{A} & =3+a_{1}\cos(2\theta_{1})\cos(2\theta_{2}),\nonumber \\
\lambda_{B} & =3-a_{1}\cos(2\theta_{1})\cos(2\theta_{2}),\nonumber \\
\lambda_{C} & =\frac{9}{16}\sin^{2}(2\theta_{1})\sin^{2}(2\theta_{2})\left[\left(c_{1}-c_{2}\right)^{2}+4c_{2}\left(\cos(\phi_{1})+\cos(\phi_{2})\right)\left(c_{2}\cos(\phi_{1})+c_{1}\cos(\phi_{2})\right)\right],\label{eq:Lambda_A-34}\\
\lambda_{1,2} & =\lambda_{B}\pm\sqrt{a_{1}^{2}\left(\cos(2\theta_{1})+\cos(2\theta_{2})\right)^{2}+\lambda_{C}},\nonumber \\
\lambda_{3,4} & =\lambda_{A}\pm\sqrt{a_{1}^{2}\left(\cos(2\theta_{1})-\cos(2\theta_{2})\right)^{2}+\lambda_{C}}.\nonumber 
\end{align}
The optimization of $S_{rel}$ is an hard task, and cannot be performed
fully analytically in a simple way. Indeed, it has been proven that
in a bipartite system the optimization of the relative entropy (for
a general density matrix) involves the solution of equations containing
logarithms of nonlinear quantities, that cannot be obtained analytically
(see for instance~\cite{GIROLAMI_ANAL_PROGR_DISC_2Q,QD_ARB_ST2}).
This is the reason why we developed a numerical approach to the minimization,
whose results have been used as guidelines to give an analytical expression
for $S_{rel}$. A similar method has already been adopted independently
to estimate the quantum discord of two-qutrit Werner states in Ref.~\cite{ANAL_EXPR_WERN_QTRIT}.

First, in our procedure, we generate randomly a suitable number of
triplets $(a_{1},c_{1},c_{2})$ (obeying to the constraints of Eq.~(\ref{eq:Constra1})), and then we minimize numerically the corresponding
expression of $S_{rel}(\theta_{1},\theta_{2},\phi_{1},\phi_{2})$
over a grid of points in the 4D-space $\mathbb{U}=R_{\theta_{1}}\times R_{\theta_{2}}\times R_{\phi_{1}}\times R_{\phi_{2}}$,
where $R_{\theta_{i}}$ and $R_{\phi_{i}}$ are the intervals $[0;\pi)$
and $[0;2\pi)$ respectively, given the periodicity of the functions
in Eqs.~(\ref{eq:Lambda_A-34}). The optimization procedure, which has been
shown to be an NP-complete problem~\cite{Yichen_QD_NPCOMPLETE}, was performed
using exhaustive enumeration (i.e. brute force search) over a grid
in the $\mathbb{U}$ space, to be sure to find the true absolute minima
of $S_{rel}$. Our calculations indicate that the function $S_{rel}$
exhibits many equivalent absolute minima, and that the ``first''
one (i.e. the one with the lowest values of its coordinates) is always
reached for $\theta_{1}=\theta_{2}=\theta$ and $\phi_{1}=0$. Specifically,
it is found alternatively in one of these three points $(\theta_{1},\theta_{2},\phi_{1},\phi_{2})$
of $\mathbb{U}$: $(0,0,0,0)$, $(\frac{\pi}{4},\frac{\pi}{4},0,0)$
or $(\frac{\pi}{4},\frac{\pi}{4},0,\bar{\phi}_{2})$, where $\bar{\phi}_{2}$
depends upon $(a_{1},c_{1},c_{2})$ %
\footnote{Notice that when $\theta_{i}=0$ other equivalent minima can be found
for $\theta_{i}=\frac{\pi}{2}$ or $\theta_{i}=\pi$, and when $\theta_{i}=\frac{\pi}{4}$
other equivalent minima can be found for $\theta_{i}=\frac{3\pi}{4}$,
but we will focus only on the cases $\theta=0$ or $\theta=\frac{\pi}{4}$,
which are the simpler ones.%
}. This means that the minimal relative entropy $S(\rho_{A|BC})$ can
take only three possible analytical forms (provided that one can find
an analytical expression for $\bar{\phi}_{2}$).

Starting from these numerical results, we performed an analytical
study on the specific case of $S_{rel}(\theta,\theta,\phi_{1},\phi_{2})$,
which confirmed that this function has two extrema in $\theta=0$
and $\theta=\frac{\pi}{4}$. Moreover, our analytical approach showed
that the function $S_{rel}(\frac{\pi}{4},\frac{\pi}{4},0,\phi_{2})$
attains its minimum value for $\sin(\phi_{2})=0$ or $\cos(\phi_{2})=\left(-\frac{c_{1}+c_{2}}{2c_{1}}\right)$,
(which holds only if certain conditions are satisfied - see Eq.~(\ref{eq:Cosine_Cond})
in the Appendix~\ref{sec:Analytical-study-of-Srel}). This is consistent
with numerical calculations, which give as minimum $\phi_{2}=0$ or
$\phi_{2}=\bar{\phi}_{2}=\arccos\left(-\frac{c_{1}+c_{2}}{2c_{1}}\right)$.
Further details are given in the Appendix~\ref{sec:Analytical-study-of-Srel}.

Our derivation leads to the following expressions for the minimum
values of $S_{rel}$:
\begin{align}
S_{1}=S_{rel}(0,0,0,0) & =1-\frac{1}{12}\gamma(a_{1}),\nonumber \\
S_{2}=S_{rel}({\textstyle \frac{\pi}{4}},{\textstyle \frac{\pi}{4}},0,0) & =1-\frac{1}{2}\varepsilon\left({\textstyle \frac{3c_{2}+c_{1}}{4}}\right),\label{eq:S1-S2-S3}\\
S_{3}=S_{rel}({\textstyle \frac{\pi}{4}},{\textstyle \frac{\pi}{4}},0,\bar{\phi}_{2}) & =1-\frac{1}{2}\varepsilon\left({\textstyle \frac{1}{4}\sqrt{\frac{(c_{1}-c_{2})^{3}}{c_{1}}}}\right),\nonumber 
\end{align}

where
\begin{equation}
\gamma(x)=(3+x)\logd(3+x)+(3-3x)\logd(3-3x)-2(3-x)\logd(3-x),\label{eq:Gamma}
\end{equation}
\begin{equation}
\varepsilon(x)=(1+x)\logd(1+x)+(1-x)\logd(1-x).\label{eq:Epsilon}
\end{equation}

When both $S_{2}$ and $S_{3}$ are well defined expressions, we found
with additional analytical calculations that $S_{3}<S_{2}$ if $c_{1}\cdot c_{2}<0$
(see Appendix~\ref{sec:Comparison-between-S2S3}). This implies that
the relative entropy takes the form:
\begin{equation}
S(\rho_{A|BC})=\begin{cases}
\min\left\{ S_{1},S_{3}\right\}  & \left|3c_{1}\right|\geq\left|c_{2}\right|\:\textrm{and}\: c_{1}\cdot c_{2}<0\\
\min\left\{ S_{1},S_{2}\right\}  & \textrm{otherwise}
\end{cases},\label{eq:Rel_entropy_final}
\end{equation}

where the minimization is required only if both entropy expressions
are well defined (considering the constraint imposed on $a_{1}$,
we can say that the expression of $S_{1}$ is always well defined,
at least in the limit given by Eq.~\ref{eq:Constra1}). In our simulations
over a set of 6000 triplets of values $(a_{1},c_{1},c_{2})$ randomly
generated, we observe that the minimum of $S(\rho_{A|BC})$ occurs
in $S_{1}$ in the $52\%$ of the cases, in $S_{2}$ in the $31\%$
of the cases and in $S_{3}$ in the remaining $17\%$ of the cases.

Finally, by using Eqs.~(\ref{eq:GenTripDisc}), (\ref{eq:S_tot}),
(\ref{eq:S_AB}) and (\ref{eq:Rel_entropy_final}), we can write the
expression for GTQD:

\begin{align}
D^{(3)}(\rho) & =S(\rho_{A|BC})-\frac{1}{6}\left(3-a_{1}\right)\logd\left(3-a_{1}\right)-\frac{1}{6}\left(3+a_{1}\right)\logd\left(3+a_{1}\right)+2+\logd(3)\nonumber \\
 & -\{3+\frac{1}{8}\left[2(3+a_{1})\log_{2}(3)-\left(1-a_{1}-c_{1}\right)\logd\left(1-a_{1}-c_{1}\right)-\left(1-a_{1}+c_{1}\right)\logd\left(1-a_{1}+c_{1}\right)\right.\nonumber \\
 & \left.-\left(3+a_{1}-3c_{2}\right)\logd\left(3+a_{1}-3c_{2}\right)-\left(3+a_{1}+3c_{2}\right)\logd\left(3+a_{1}+3c_{2}\right)\right]\}.
\end{align}

\section{RESULTS AND DISCUSSION\label{sec:RESULTS-AND-DISCUSSION}}

To validate our approach, we apply the above expression to two prototypical
cases of study. In particular, the well known result $D^{(3)}(\rho_{GHZ})=1$
for a pure GHZ state $\rho_{GHZ}=\left|GHZ\right\rangle \left\langle GHZ\right|=\frac{1}{2}(\left|000\right\rangle +\left|111\right\rangle )(\left\langle 000\right|+\left\langle 111\right|)$,
is obtained by setting $a_{1}=-3$, $c_{1}=\pm4$ and $c_{2}=0$ in
Eq.~(\ref{eq:Rho_X}).

Analogously, it can be shown that for a maximally mixed state with
$c_{1}=c_{2}=0$ we find $D^{(3)}(\rho)=0$ (and $S(\rho_{A|BC})=S(\rho)-S(\rho_{A,B})=S_{1}$,
since in this case $S_{2}=S_{3}=1$) whatever the value of $a_{1}$
is, as expected since all correlations are classical.

\begin{figure}
\begin{minipage}[c]{.48\textwidth}
\begin{centering}
\includegraphics[scale=0.47]{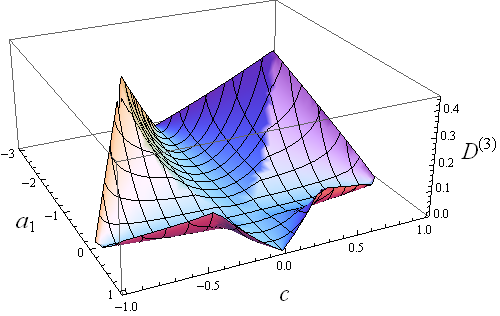}
\par\end{centering}
\caption{Genuine Tripartite Quantum Discord for $c_{1}=c_{2}=c$ (notice that
the maximum value of \emph{z} axis is set to 0.45 and not to 1.0 in
order to make the graph more readable).}
\label{GTDiscord(c1=00003Dc2)}
\end{minipage}
\quad\; \begin{minipage}[c]{.48\textwidth}
\begin{centering}
\includegraphics[scale=0.47]{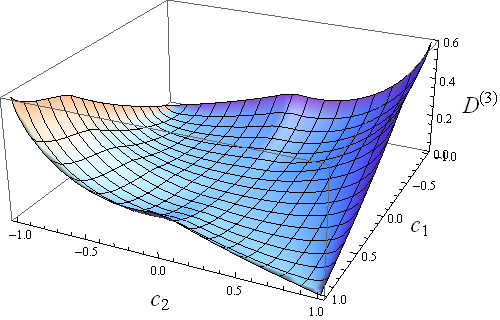}
\par\end{centering}
\caption{Genuine Tripartite Quantum Discord for $a_{1}=0$ (notice that the
maximum value of \emph{z} axis is set to 0.6 and not to 1.0 in order
to make the graph more readable).}
\label{GTDiscord(a1=00003D0)}
\end{minipage}
\end{figure}

Moving towards a more general case, we can set $c_{1}=c_{2}=c$ and
plot the values of $D^{(3)}(\rho)$ with respect to $a_{1}$ and $c$.
As we see from Figure~\ref{GTDiscord(c1=00003Dc2)}, along the line
$c=0$ (maximally mixed states) the genuine tripartite discord vanishes
- as explained above. Moreover, $D^{(3)}(\rho)$ is zero also along
the line $a_{1}=0$, which does not corresponds to mixed states, but
to a case where again we have $S(\rho_{A|BC})=S(\rho)-S(\rho_{A,B})=S_{2}$
(since here $S_{1}=S_{3}=1$). Maximum values of $D^{(3)}(\rho)$
are achieved when $a_{1}\simeq1.38$ and $c\simeq0.54$.

A further analysis of the states with $a_{1}=0$ is performed by investigating
the behavior of GTQD as a function of $c_{1}$ and $c_{2}$ (see
Figure~\ref{GTDiscord(a1=00003D0)}). When $c_{1}\neq c_{2}$ the
GTQD never goes to zero, and it reaches its maximum value in $(c_{1},c_{2})=(1,-1)$
or $(-1,1)$, where $D^{(3)}(\rho)=1-\frac{1}{2}\varepsilon\left(\frac{1}{\sqrt{2}}\right)$.
In detail, the density matrix $\rho'$ obtained by setting $(c_{1},c_{2})=(1,-1)$
in Eq.~(\ref{eq:Rho_X}) is a linear combination of density matrices
of pure GHZ states of the type:

\begin{equation}
\rho_{GHZ}^{\pm}(k)=\frac{1}{2}\left(\left|k\right\rangle \pm\left|\bar{k}\right\rangle \right)\left(\left\langle k\right|\pm\left\langle \bar{k}\right|\right),
\end{equation}

where $k$ is a three bit binary number (from $0$ to $7$) and $\bar{k}$
is the result of flipping each bit of $k$~\cite{WEINSTEIN_ENTNGL_3Q_X}.
Indeed:

\begin{equation}
\rho'=\frac{1}{4}\left(\rho_{GHZ}^{+}(0)+\rho_{GHZ}^{-}(1)+\rho_{GHZ}^{-}(2)+\rho_{GHZ}^{-}(3)\right),
\end{equation}

and a similar expression can be found for $\rho$ when $(c_{1},c_{2})=(-1,1)$.
Unlike pure GHZ states, this mixed state $\rho'$ is not a maximally
entangled one (indeed its negativity is zero, as we will see in the
following), but it shows a GTQD different from zero. Moreover, also
the state $\rho''$ obtained by setting $(c_{1},c_{2})=(1,1)$ is
a linear combination of pure GHZ states:

\begin{equation}
\rho''=\frac{1}{4}\left(\rho_{GHZ}^{+}(0)+\rho_{GHZ}^{+}(1)+\rho_{GHZ}^{+}(2)+\rho_{GHZ}^{+}(3)\right),
\end{equation}

but this state is characterized by zero GTQD. A similar expression
can be found for $\rho$ when $(c_{1},c_{2})=(-1,-1)$, and the value
of GTQD is again zero. Therefore, we conclude that a linear combinations
of GHZ states is characterized by zero discord when all the states
are of kind $\rho_{GHZ}^{+}(k)$ (or $\rho_{GHZ}^{-}(k)$), as it
occurs for bipartite systems when we combine linearly Bell states
with the same sign. Otherwise, if we combine GHZ states of kind $\rho_{GHZ}^{+}(k)$
and $\rho_{GHZ}^{-}(k)$ together, the GTQD can be different from
zero.

\begin{figure}
\begin{minipage}[c]{.48\textwidth}
\begin{centering}
\includegraphics[scale=0.47]{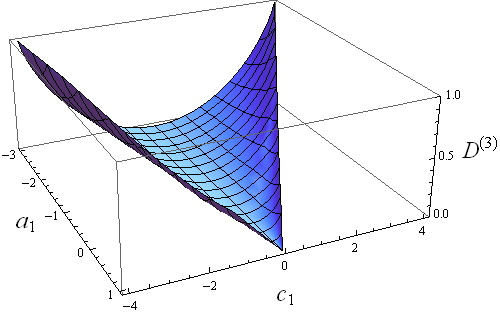}
\par\end{centering}
\caption{Genuine Tripartite Quantum Discord for $c_{2}=0$.}
\label{GTDiscord(c2=00003D0)}
\end{minipage}
\quad\; \begin{minipage}[c]{.48\textwidth}
\begin{centering}
\includegraphics[scale=0.47]{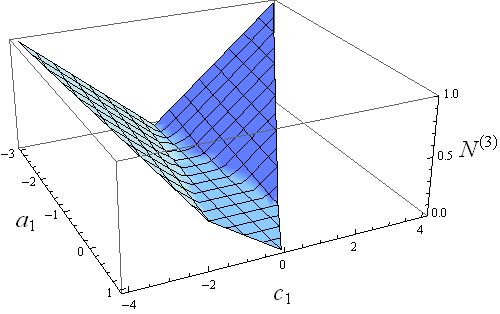}
\par\end{centering}
\caption{Tripartite Negativity for $c_{2}=0$.}
\label{TNeg(c2=00003D0)}
\end{minipage}
\end{figure}

Finally, we study $D^{(3)}(\rho)$ by setting $c_{2}=0$. We see in
Figure~\ref{GTDiscord(c2=00003D0)} that the GTQD vanishes along the
line $c_{1}=0$ and reaches its absolute maximum value (as expected)
for the maximally entangled GHZ states $(a_{1},c_{1})=(3,\pm4)$.

Now we compare GTQD and tripartite entanglement, where the latter
is quantified by\emph{ }means of tripartite negativity $N^{(3)}(\rho)$,
given in Eq.~(\ref{eq:trip_neg}). For the state of Eq.~(\ref{eq:Rho_X})
we get:
\begin{align}
N^{(3)}(\rho)=\frac{1}{24}\left(\left|3+a_{1}-3c_{1}\right|+\left|3+a_{1}+3c_{1}\right|+\right.\nonumber \\
+\left.2\left|3+a_{1}-3c_{2}\right|+2\left|3+a_{1}+3c_{2}\right|+
3\left|1-a_{1}-c_{2}\right|+3\left|1-a_{1}+c_{2}\right|\right)-1.
\end{align}

By evaluating this expression for some special values of the parameters
$(a_{1},c_{1},c_{2})$, we see that there are regions in which the
tripartite entanglement cannot be detected (i.e. the negativity is
zero) but on the other hand the GTQD is different from zero. In particular,
for $c_{1}=c_{2}$ or $a_{1}=0$ the tripartite negativity is zero
everywhere. On the contrary, for $c_{2}=0$ (see Figure~\ref{TNeg(c2=00003D0)})
there are regions where GTQD can be both smaller or larger than the
negativity. This result is not surprising. Indeed, in bipartite systems
for some cases entanglement has been found to be larger than quantum
discord~\cite{LUO_MEAS,ANAL_QD_2QUBIT_X_states}, since the latter
cannot simply be considered as the sum of the entanglement and other
forms of nonclassical correlations~\cite{ANAL_QD_2QUBIT_X_states}.
However, in our case we can explain this result also on the grounds
that tripartite negativity quantifies a tripartite entanglement not
necessarily genuine~\cite{Sabin_Trip_Negat,EXP_MULTIP_SPINCHAIN}, so it
could detect in principle a larger amount of quantum correlations
with respect to a genuine quantifier, such as GTQD.

\section{CONCLUSIONS\label{sec:CONCLUSIONS}}

In this paper, we have developed a hybrid analytical-numerical approach
to find the analytical expression for GTQD $D^{(3)}(\rho)$, specifically
for a subclass of X-states, symmetrical under exchange and flip of all 
qubits, which are defined by three parameters ($a_{1}$, $c_{1}$ and $c_{2}$). The expression of $D^{(3)}(\rho)$
depends on the relative entropy $S(\rho_{A|BC})$, whose estimation
requires the minimization of the function $S_{rel}(\theta_{1},\theta_{2},\phi_{1},\phi_{2})$
depending on 4 angular variables. Numerical calculations show that
$S_{rel}$ possesses only three different minimum points, which should
correspond to three distinct analytical expressions for $S(\rho_{A|BC})$.
Further analytical studies performed over a simplified form of $S_{rel}(\theta_{1},\theta_{2},\phi_{1},\phi_{2})$
allowed us to find the exact analytical expressions for $S(\rho_{A|BC})$
and the conditions under which they can be used.
These analytical findings have been compared with some thousand of
numerical simulations and have been proven always right. Moreover,
they are able to reproduce the known results for GTQD in some simple
systems, namely GHZ and maximally mixed states. When confronted with
tripartite negativity, the calculations show that there are regions
in the space of the parameters ($a_{1}$, $c_{1}$ and $c_{2}$) where
entanglement cannot be detected, while genuine quantum correlations
(evaluated in terms of GTDQ) differ from zero.

Further possible development of this work include the analytical study
of the time evolution of genuine quantum correlations (as accounted
by GTQD) and the extension of the hybrid approach here developed to
more general cases.

\appendix

\section{Appendix: Relative Entropies Definition\label{sec:Relative-Entropies-Definition}}

Following \emph{Zhao et al.}~\cite{ZHAO_GENUIN_TRIP}, we can define
the \emph{relative entropy} $S(\rho_{A|BC})$ for tripartite systems
as:
\begin{equation}
S(\rho_{A|BC})=\underset{\{E_{ij}^{BC}\}}{\min}\sum_{ij}p_{ij}S(\rho_{A|E_{ij}^{BC}}),\label{eq:EntrRelBC}
\end{equation}

where:

\begin{equation}
\rho_{A|E_{ij}^{BC}}=\frac{\tilde{\rho}_{A|E_{ij}^{BC}}}{p_{ij}}=\frac{1}{p_{ij}}\Tr_{B,C}\left[\left(I^{A}\otimes E_{ij}^{BC}\right)\rho\right],\label{eq:RhoA|EBC}
\end{equation}
\begin{equation}
p_{i,j}=\Tr_{A,B,C}\left[\left(I^{A}\otimes E_{ij}^{BC}\right)\rho\right].\label{eq:pij_RhoA|EBC}
\end{equation}

In the previous expressions, the operators $E_{ij}^{BC}$ are \emph{positive-operator-valued
measures} (POVMs) that act on parties $B$ and $C$ (i.e. in the Hilbert
space $\mathcal{H}_{BC}=\mathcal{H}_{B}\otimes\mathcal{H}_{C}$),
and whose outcomes are labeled with two indices ($i,j$). For sake
of simplicity, we will replace the global POVM $E_{ij}^{BC}$ with
the external product of two local POVMs, acting separately on parties
$B$ and $C$, using the same procedure given in~\cite{GIROLAMI_ANAL_PROGR_DISC_2Q}.
Moreover, following the convention in literature~\cite{H_VEDR,ZUR_DISC,HAMIEH_POVMs,GIORGI_GENUIN_MULTIP,ZHAO_GENUIN_TRIP}),
we use orthogonal \emph{projection-valued measures} (PVMs) to optimize
entropy in Eq.~(\ref{eq:EntrRelBC}), since they are easier to implement
in the numerical minimization process %
\footnote{This approach has been recently questioned by \emph{Zhao et al.}:
in their paper~\cite{ZHAO_GENUIN_TRIP}, they show that the product
POVM $E_{i}^{B}\otimes E_{j}^{C}$ may not be the optimal POVM $E_{ij}^{BC}$
that minimizes genuine tripartite discord. However, we must notice
that the qualitative behaviors of $D^{(3)}(\rho)$ are not changed
by this approach (except for the overestimation of $D^{(3)}(\rho)$),
i.e. both approaches are able to record the presence of GTQD and its
increasing (or decreasing) trend, according to the variations of the
parameters which define the density operator $\rho$.%
}. Then, the measurement operators are:
\begin{align}
E_{ij}^{BC} & \rightarrow\Pi_{i}^{B}\otimes\Pi_{j}^{C}=\left|\beta_{i}\right\rangle \left\langle \beta_{i}\right|\otimes\left|\gamma_{j}\right\rangle \left\langle \gamma_{j}\right|,\label{eq:Eqproj}
\end{align}
where $\left|\beta_{i}\right\rangle $ and $\left|\gamma_{j}\right\rangle $
are orthogonal normalized basis states of the Hilbert spaces $\mathcal{H}_{B}$
and $\mathcal{H}_{C}$, respectively.

A possible parametrization of the basis vectors $\left|\beta_{i}\right\rangle $
and $\left|\gamma_{j}\right\rangle $ with respect to the standard
basis $\{\left|0\right\rangle ,\left|1\right\rangle \}$ can be found
in literature (see \cite{FABRIZIO_GENUIN}; for a full derivation
of the basis vectors see \cite{ANAL_EXPR_WERN_QTRIT}):
\begin{align}
\left|\beta_{1}\right\rangle  & =\cos\theta_{1}\left|0_{B}\right\rangle +e^{+i\phi_{1}}\sin\theta_{1}\left|1_{B}\right\rangle ,\label{eq:vectorb1}\\
\left|\beta_{2}\right\rangle  & =\sin\theta_{1}\left|0_{B}\right\rangle -e^{+i\phi_{1}}\cos\theta_{1}\left|1_{B}\right\rangle ,\label{eq:vectorb2}\\
\left|\gamma_{1}\right\rangle  & =\cos\theta_{2}\left|0_{C}\right\rangle +e^{+i\phi_{2}}\sin\theta_{2}\left|1_{C}\right\rangle ,\label{eq:vectorc1}\\
\left|\gamma_{2}\right\rangle  & =\sin\theta_{2}\left|0_{C}\right\rangle -e^{+i\phi_{2}}\cos\theta_{2}\left|1_{C}\right\rangle ,\label{eq:vectorc2}
\end{align}

where the angles $\theta_{i}$ and $\phi_{i}$ belong to the interval
$[0;2\pi)$.

Since we are studying a system whose state is symmetrical under any
permutation of its subsystems, any subscript or superscript referring
to a particular subsystem in the relative entropy expression (\ref{eq:EntrRelBC})
can be dropped. Now, recalling the sum rule $\sum_{l=1}^{2}\tilde{\lambda}_{l}^{(ij)}=p_{ij}$
for the eigenvalues $\tilde{\lambda}_{l}^{(ij)}$ of $\tilde{\rho}_{ij}=\tilde{\rho}_{A|E_{ij}^{BC}}$
(crf. Eqs.~(\ref{eq:RhoA|EBC}) and (\ref{eq:pij_RhoA|EBC})), we
can simplify Eq. (\ref{eq:EntrRelBC}) as follows:
\begin{equation}
S(\rho_{A|BC})=\underset{\{E_{ij}^{BC}\}}{\min}\left[-H(p)+\sum_{i,j}S(\tilde{\rho}_{ij})\right],\label{eq:simplifSrel}
\end{equation}

where $H(p)=-\sum_{i,j}p_{ij}\logd(p_{ij})$ is the Shannon Entropy
of the probability ensemble$\{p_{ij}\}$.

Now, using Eqs.~(\ref{eq:vectorb1})-(\ref{eq:vectorc2}) to write
the PVMs - together with the change of variables ($\phi_{1}-\phi_{2}\rightarrow\phi_{1}$,
$\phi_{1}+\phi_{2}\rightarrow\phi_{2}$), which simplifies our calculations
- the relative entropy in (\ref{eq:simplifSrel}) can be written as
a function of four angular variables:
\begin{align}
S(\rho_{A|BC}) & =\underset{\theta_{i},\phi_{i}}{\min}\, S_{rel}(\theta_{1},\theta_{2},\phi_{1},\phi_{2}).\label{eq:SrelOptim}
\end{align}

The final expression for $S(\rho_{A|BC})$, with all terms written
explicitly, is given in Section~\ref{sec:Rel_Entr_Minim}.

\section{Appendix: Analytical study of $S_{rel}(\theta_{1},\theta_{2},\phi_{1},\phi_{2})$\label{sec:Analytical-study-of-Srel}}

The relative entropy $S_{rel}(\theta_{1},\theta_{2},\phi_{1},\phi_{2})$
of Eq.~(\ref{eq:SA|BC}) 
\begin{equation}
S_{rel}(\theta_{1},\theta_{2},\phi_{1},\phi_{2})=1+\frac{1}{6}\left[\lambda_{A}\logd\lambda_{A}+\lambda_{B}\logd\lambda_{B}\right]-\frac{1}{12}\sum_{i=1}^{4}\lambda_{i}\logd\lambda_{i}
\end{equation}

can be studied in a simplified form setting $\theta_{1}=\theta_{2}=\theta$.
Under this condition, the $\lambda_{j}$ of Eqs.~(\ref{eq:Lambda_A-34})
become:
\begin{align}
\lambda_{A} & =3+a_{1}\cos^{2}(2\theta),\nonumber \\
\lambda_{B} & =3-a_{1}\cos^{2}(2\theta),\nonumber \\
\lambda_{C} & =\frac{9}{16}\sin^{4}(2\theta)\, f(\phi_{1},\phi_{2}),\\
f(\phi_{1},\phi_{2}) & =\left[\left(c_{1}-c_{2}\right)^{2}+4c_{2}\left(\cos(\phi_{1})+\cos(\phi_{2})\right)\left(c_{2}\cos(\phi_{1})+c_{1}\cos(\phi_{2})\right)\right]\nonumber \\
\lambda_{1,2} & =\lambda_{B}\pm\sqrt{4a_{1}^{2}\cos^{2}(2\theta)+\lambda_{C}},\nonumber \\
\lambda_{3,4} & =\lambda_{A}\pm\sqrt{\lambda_{C}}.\nonumber 
\end{align}

The minima of $S_{rel}$ must satisfy the equation:
\begin{equation}
\frac{\partial S_{rel}(\theta,\theta,\phi_{1},\phi_{2})}{\partial\theta}=0,
\end{equation}

which can be rewritten as follows:
\begin{equation}
\frac{\partial S_{rel}}{\partial\lambda_{A}}\frac{\partial\lambda_{A}}{\partial\theta}+\frac{\partial S_{rel}}{\partial\lambda_{B}}\frac{\partial\lambda_{B}}{\partial\theta}+\sum_{i=1}^{4}\frac{\partial S_{rel}}{\partial\lambda_{i}}\frac{\partial\lambda_{i}}{\partial\theta}=0.\label{eq:Srel_DerTheta_cmpct}
\end{equation}

The derivatives of the $\lambda_{j}$ appearing in Eq.~(\ref{eq:Srel_DerTheta_cmpct})
are given by:
\begin{align}
\frac{\partial\lambda_{A}}{\partial\theta} & =-4a_{1}\cos(2\theta)\sin(2\theta)\nonumber \\
\frac{\partial\lambda_{B}}{\partial\theta} & =+4a_{1}\cos(2\theta)\sin(2\theta)\nonumber \\
\frac{\partial\lambda_{C}}{\partial\theta} & =\frac{9}{2}\sin^{3}(2\theta)\cos(2\theta)f(\phi_{1},\phi_{2})\\
\frac{\partial\lambda_{1,2}}{\partial\theta} & =\frac{\partial\lambda_{B}}{\partial\theta}\pm\frac{16a_{1}^{2}\cos(2\theta)\sin(2\theta)+\frac{\partial\lambda_{C}}{\partial\theta}}{2\sqrt{4a_{1}^{2}\cos^{2}(2\theta)+\lambda_{C}},},\nonumber \\
\frac{\partial\lambda_{3,4}}{\partial\theta} & =\frac{\partial\lambda_{A}}{\partial\theta}\pm\frac{\frac{\partial\lambda_{C}}{\partial\theta}}{2\sqrt{\lambda_{C}},}\nonumber 
\end{align}

and furthermore:
\begin{align}
\frac{\partial S_{rel}}{\partial\lambda_{A,B}} & =+\frac{1}{6\ln2}(\ln\lambda_{A,B}+1)\\
\frac{\partial S_{rel}}{\partial\lambda_{i}} & =-\frac{1}{12\ln2}(\ln\lambda_{i}+1)\quad(i=1,2,3,4)
\end{align}

Eq.~(\ref{eq:Srel_DerTheta_cmpct}) then becomes:
\begin{multline}
\cos(2\theta)\sin(2\theta)\left[-4a_{1}\frac{\partial S_{rel}}{\partial\lambda_{A}}+4a_{1}\frac{\partial S_{rel}}{\partial\lambda_{B}}+\right.\\
\left(4a_{1}+\frac{16a_{1}^{2}+\frac{9}{2}\sin^{2}(2\theta)f(\phi_{1},\phi_{2})}{2\sqrt{4a_{1}^{2}\cos^{2}(2\theta)+\lambda_{C}},}\right)\frac{\partial S_{rel}}{\partial\lambda_{1}}+\left(4a_{1}-\frac{16a_{1}^{2}+\frac{9}{2}\sin^{2}(2\theta)f(\phi_{1},\phi_{2})}{2\sqrt{4a_{1}^{2}\cos^{2}(2\theta)+\lambda_{C}},}\right)\frac{\partial S_{rel}}{\partial\lambda_{2}}+\\
\left.\left(-4a_{1}+\frac{\frac{9}{2}\sin^{2}(2\theta)f(\phi_{1},\phi_{2})}{2\sqrt{\lambda_{C}},}\right)\frac{\partial S_{rel}}{\partial\lambda_{3}}+\left(-4a_{1}-\frac{\frac{9}{2}\sin^{2}(2\theta)f(\phi_{1},\phi_{2})}{2\sqrt{\lambda_{C}},}\right)\frac{\partial S_{rel}}{\partial\lambda_{4}}\right]=0\label{eq:Srel_DerTheta_expand}
\end{multline}

This expression shows that $\frac{\partial S_{rel}}{\partial\theta}=0$
when $\sin(2\theta)=0$ or $\cos(2\theta)=0$, that is the function
can attain its minimum value for a value in the set $\theta=0+k\frac{\pi}{2}$
or $\theta=\frac{\pi}{4}+k\frac{\pi}{2}$, where $k\in\mathbb{Z}$.
Indeed, it can be shown that the whole l.h.s. of Eq.~(\ref{eq:Srel_DerTheta_expand})
goes to zero when $\theta$ approaches in the limit the values listed
before.

When we make the further assumption that $\theta=\frac{\pi}{4}$ and
$\phi_{1}=0$ (as suggested by numerical calculations), we get:
\begin{align}
\lambda_{A} & =3,\nonumber \\
\lambda_{B} & =3,\nonumber \\
\lambda_{C} & =\frac{9}{16}\, f(0,\phi_{2}),\\
f(0,\phi_{2}) & =\left[\left(c_{1}-c_{2}\right)^{2}+4c_{2}\left(1+\cos(\phi_{2})\right)\left(c_{2}+c_{1}\cos(\phi_{2})\right)\right],\nonumber \\
\lambda_{1,2} & =\lambda_{3,4}=3\pm\sqrt{\lambda_{C}},\nonumber 
\end{align}

and
\begin{equation}
S_{rel}({\textstyle \frac{\pi}{4}},{\textstyle \frac{\pi}{4}},0,\phi_{2})=1+\logd3-\frac{1}{6}\sum_{i=1}^{2}\lambda_{i}\logd\lambda_{i}.
\end{equation}

Therefore, the minimum is reached when
\begin{equation}
\frac{\partial S_{rel}({\textstyle \frac{\pi}{4}},{\textstyle \frac{\pi}{4}},0,\phi_{2})}{\partial\phi_{2}}=0,
\end{equation}

that is
\begin{equation}
\frac{\partial}{\partial\phi_{2}}\left(\lambda_{1}\logd\lambda_{1}+\lambda_{2}\logd\lambda_{2}\right)=0.
\end{equation}

With further simplifications we get:
\begin{equation}
\left[\ln\frac{\left(3+\sqrt{\lambda_{C}}\right)}{\left(3-\sqrt{\lambda_{C}}\right)}\frac{1}{2\ln2\sqrt{\lambda_{C}}}\right]\frac{\partial\lambda_{C}}{\partial\phi_{2}}=0.
\end{equation}

The expression in the square brackets is always greater than 0, since
the argument of the logarithm is always greater than 1 if $\lambda_{C}>0$,
and when $\lambda_{C}\rightarrow0$ the limit is finite, positive
and different from zero. Therefore the extremum can be found only
for:
\begin{equation}
\frac{\partial\lambda_{C}}{\partial\phi_{2}}=0\quad\Longleftrightarrow\quad\frac{\partial f(0,\phi_{2})}{\partial\phi_{2}}=0,
\end{equation}

which leads to the final equation:
\begin{equation}
\sin(\phi_{2})\left(c_{2}+c_{1}+2c_{1}\cos(\phi_{2})\right)=0.
\end{equation}

The solutions are:
\begin{equation}
\sin(\phi_{2})=0\quad\text{or}\quad\cos(\phi_{2})=\left(-\frac{c_{1}+c_{2}}{2c_{1}}\right),
\end{equation}

that is
\begin{equation}
\phi_{2}=0+k\pi\quad\text{or}\quad\phi_{2}=\pm\arccos\left(-\frac{c_{1}+c_{2}}{2c_{1}}\right)+2k\pi,
\end{equation}

where $k\in\mathbb{Z}$. Clearly, the second set of extrema exists
only if:
\begin{equation}
-1\leq-\frac{c_{1}+c_{2}}{2c_{1}}\leq+1\quad\Rightarrow\quad\begin{cases}
c_{1}>0\quad\textrm{and}\quad & -3c_{1}\leq c_{2}\leq c_{1}\\
c_{1}<0\quad\textrm{and}\quad & c_{1}\leq c_{2}\leq-3c_{1}
\end{cases}.\label{eq:Cosine_Cond}
\end{equation}

\section{Appendix: Comparison between $S_{2}$ and $S_{3}$\label{sec:Comparison-between-S2S3}}

When $\theta={\textstyle \frac{\pi}{4}}$ and $\phi_{1}=0$, the expression
for $S_{rel}$ can be written as:
\begin{align}
S_{rel}({\textstyle \frac{\pi}{4}},{\textstyle \frac{\pi}{4}},0,\phi_{2}) & =1+\logd3-\frac{1}{6}\left[\left(3+\sqrt{\lambda_{C}}\right)\logd\left(3+\sqrt{\lambda_{C}}\right)+\left(3-\sqrt{\lambda_{C}}\right)\logd\left(3-\sqrt{\lambda_{C}}\right)\right]\nonumber \\
 & =1-\frac{1}{2}\left[\left(1+\frac{1}{3}\sqrt{\lambda_{C}}\right)\logd\left(1+\frac{1}{3}\sqrt{\lambda_{C}}\right)+\left(1-\frac{1}{3}\sqrt{\lambda_{C}}\right)\logd\left(1-\frac{1}{3}\sqrt{\lambda_{C}}\right)\right]\nonumber \\
 & =1-\frac{1}{2}\varepsilon\left(\frac{1}{3}\sqrt{\lambda_{C}}\right),
\end{align}

where $\varepsilon(x)$ is given by (\ref{eq:Epsilon}). The function
$S(x)=1-\frac{1}{2}\varepsilon(x)$ is known in the literature as
an estimator of correlations and relative entropies in bipartite systems~\cite{LUO_QBIT}, and its expression holds only for $-1\leq x\leq1$
(in our case it is always $x>0$). Due to its symmetry properties,
$S(x)$ has its maximum value for $x=0$, and decreases monotonically
as $x$ approaches $1$ (or $-1$). Therefore we conclude that:
\begin{equation}
S(x_{1})<S(x_{2})\Longleftrightarrow x_{1}>x_{2}\quad\forall x_{1},x_{2}\geq0\label{eq:Condition_S(x)}
\end{equation}

If $\phi_{2}=0$, then $\lambda_{C}$ takes the following value:
\begin{equation}
\lambda_{C}=\frac{9}{16}\, f(0,0)=\frac{9}{16}(3c_{2}+c_{1})^{2},
\end{equation}

and the corresponding expression for $S_{rel}$ is:
\begin{equation}
S_{2}=S_{rel}({\textstyle \frac{\pi}{4}},{\textstyle \frac{\pi}{4}},0,0)=1-\frac{1}{2}\varepsilon\left(\frac{|3c_{2}+c_{1}|}{4}\right)=1-\frac{1}{2}\varepsilon\left(\frac{3c_{2}+c_{1}}{4}\right)
\end{equation}

The $\lambda_{C}$ expression for $\phi_{2}=\bar{\phi}_{2}=\arccos\left(-\frac{c_{1}+c_{2}}{2c_{1}}\right)$
is the following one:
\begin{equation}
\lambda_{C}=\frac{9}{16}\, f(0,\bar{\phi}_{2})=\frac{9}{16}\frac{\left(c_{1}-c_{2}\right)^{3}}{c_{1}},
\end{equation}

which appears under a square root (for a real eigenvalue), and therefore
is acceptable only if:
\begin{equation}
\frac{\left(c_{1}-c_{2}\right)}{c_{1}}\geq0\quad\Rightarrow\quad\begin{cases}
c_{1}\geq c_{2} & \textrm{if}\; c_{1}>0\\
c_{1}\leq c_{2}. & \textrm{if}\; c_{1}<0
\end{cases}.\label{eq:Exist_LambdaC3}
\end{equation}

The corresponding expression for $S_{rel}$ becomes:
\begin{equation}
S_{3}=S_{rel}({\textstyle \frac{\pi}{4}},{\textstyle \frac{\pi}{4}},0,\bar{\phi}_{2})=1-\frac{1}{2}\varepsilon\left({\textstyle \frac{1}{4}\sqrt{\frac{\left(c_{1}-c_{2}\right)^{3}}{c_{1}}}}\right)
\end{equation}

When both $S_{2}$ and $S_{3}$ expressions are well defined (see
Eq. (\ref{eq:S1-S2-S3})), then the absolute minimum of $S_{rel}$
can occur only in the lowest of these two values, and according to
Eq. (\ref{eq:Condition_S(x)}) we conclude that
\begin{equation}
S_{3}<S_{2}\Longleftrightarrow\frac{\left(c_{1}-c_{2}\right)^{3}}{c_{1}}>(3c_{2}+c_{1})^{2}\Longleftrightarrow c_{1}\cdot c_{2}<0.\label{eq:S2S3condition}
\end{equation}

Now, collecting together the last Eq.~(\ref{eq:S2S3condition}) and
the existence conditions for $S_{3}$ (Eqs.~(\ref{eq:Cosine_Cond})
and (\ref{eq:Exist_LambdaC3})) we conclude that:
\begin{equation}
S_{3}<S_{2}\Longleftrightarrow c_{1}\cdot c_{2}<0\quad\textrm{and}\quad|c_{2}|\leq|3c_{1}|.
\end{equation}

\bibliographystyle{spphys}       	




\end{document}